\documentclass[%
 preprint,
 amsmath,amssymb,
 aps, physrev,
]{revtex4-2}

\usepackage{graphicx}
\usepackage{dcolumn}
\usepackage{bm}
\usepackage{hyperref}
\usepackage{academicons}
\usepackage{xcolor}
\usepackage{orcidlink}

\newcommand{\orcid}[1]{\href{https://orcid.org/#1}{\textcolor[HTML]{A6CE39}{\aiOrcid}}}

\begin{document}

\preprint{APS/123-QED}

\title{\textbf{Low Reynolds number pumping near an elastic half space} 
}%

\author{Avery Trevino\orcidlink{0000-0003-3481-1437},$^{1,2}$
  Thomas R. Powers\orcidlink{0000-0003-3432-8226},$^{1,2,3,4}$
  Roberto Zenit\orcidlink{0000-0002-2717-4954},$^{1,2}$
 and Mauro Rodriguez Jr.\orcidlink{0000-0003-0545-0265}$^{1,2}$}\email{Contact author: mauro\_rodriguez@brown.edu}

\affiliation{$^{1}$School of Engineering, Brown University,
Providence, RI 02912, USA \\
$^{2}$Center for Fluid Mechanics, Brown University,
Providence, RI 02912, USA\\
$^{3}$Department of Physics, Brown University,
Providence, RI 02912, USA\\
$^{4}$Brown Center for Theoretical Physics, Brown University,
Providence, RI 02912, USA}

\date{\today}

\begin{abstract}
Previous studies on peristalsis, the pumping of fluid along a channel by wave-like displacements of the channel walls, have shown that the elastic properties of the channel and the peristaltic wave shape can influence the flow rate. Motivated by the oscillatory flow of cerebrospinal fluid along compliant perivascular spaces, we consider a prescribed wave motion of a single boundary which pumps fluid at small Reynolds number near an elastic half space. We investigate the relationship between flow rate and elastic deformation as a function of the fluid and solid properties. We consider transverse and longitudinal motion of the driving peristaltic wave. We find that a transverse peristaltic wave produces net forward flow and induces elastic motion in which all material points oscillate uniformly. Conversely, a longitudinal peristaltic wave produces a net backward flow and drives elastic motion which is nonuniform in the elastic solid. We use dimensional values relevant to the flow of cerebrospinal fluid and find agreement with \textit{in vivo} velocity data.
\end{abstract}

\maketitle

\section{Introduction}

Peristalsis is a key mechanism of fluid propulsion in biological and mechanical systems, such as in the digestive system, ureter, glymphatic system~\cite{shapiro_peristaltic_1969, takagi_peristaltic_2011,romano_peristaltic_2020,bauerle_living_2020,hadaczek_perivascular_2006, wang_fluid_2011}, and some manufactured pumps \cite{forouzandeh_review_2021}. In peristalsis, a fluid is transported through a channel as a result of periodic displacements of the boundary which travel along the channel as a wave. In biological systems, this displacement typically results from synchronized contraction along a fluid-filled tube. In engineered devices the boundary deflection is often achieved with rollers or sliders which glide along the outer surface of a flexible channel. In a mechanically similar problem, the flow over undulating sheets has been studied to understand the swimming of microscopic organisms infinitely far from and near rigid and deformable boundaries \cite{taylor_1951,katz_propulsion_1974,reynolds_swimming_1965,dias_swimming_2013,shaik_swimming_2019,leshansky_enhanced_2009,pandey_optimal_2023,tchoufag_2019,jha2024taylorsswimmingsheetnear}. In these studies, rather than being fixed as in peristalsis, the sheet is free to translate due to its wave-like motion, and thus swims as in microscopic locomotion. 

Early models of peristalsis were inspired by biological flows, particularly flow which occurs in the ureter \cite{weinberg_experimental_1971,floryan_peristaltic_2021,jaffrin_peristaltic_1971,fung_peristaltic_1968}. These models were developed for low Reynolds number flows driven by small amplitude, long wavelength peristaltic profiles. Accurate biological models required the consideration of the interaction of fluid flow with the soft tissue of the channel, such as the ureter wall. The authors of \cite{bauerle_living_2020} found that naturally occurring peristaltic pumping found in network-forming slime mold \textit{Physarum polycephalum} optimally combines transverse peristaltic wave modes to maximize flow rate. While it has been argued that a longitudinal component of the peristaltic wave form can lead to reflux, or back flow, within the fluid channel \cite{katz_propulsion_1974,reynolds_swimming_1965,Blake_1971,shaik_swimming_2019}, the authors of
\cite{kalayeh_longitudinal_2023} found the opposite. In this work, to manipulate the resultant wave shape, we consider linear combinations of transverse and longitudinal wave modes. 

Recent research in the flow of cerebrospinal fluid (CSF) has led to the modelling of a peristaltic mechanism named perivascular pumping, in which the pulsatile deformation of blood vessels transports surrounding fluid through annular spaces between the artery and brain \cite{romano_peristaltic_2020,hadaczek_perivascular_2006,thomas_fluid_2019,wang_fluid_2011,Mestre2018,Mestre2020,kelley_cerebrospinal_nodate}. The softness and elasticity of the surrounding brain tissue is hypothesized to affect the flow rate of CSF through these perivascular spaces. This instance of peristaltic flow is of particular medical interest because of its relevance to the healthy operation of the gylmphatic system, the brain's system of mechanisms which coordinate in order to clear metabolic waste from extracellular tissue. The accumulation of metabolic waste such as amyloid-$\beta$ and $\tau$ proteins has been linked to neurodegenerative diseases such as Alzheimer's disease and dementia \citep{nedergaard_glymphatic_2020,brinker_new_2014,astara_novel_2023,carlstrom_clinical_2022}. Therefore, a fundamental understanding of the interaction between arterial deformations, CSF flow and brain tissue mechanics is necessary to describe glymphatic flow and its inhibitors. 
 \begin{figure*}[h]
    \centering
    \includegraphics[width=1\linewidth]{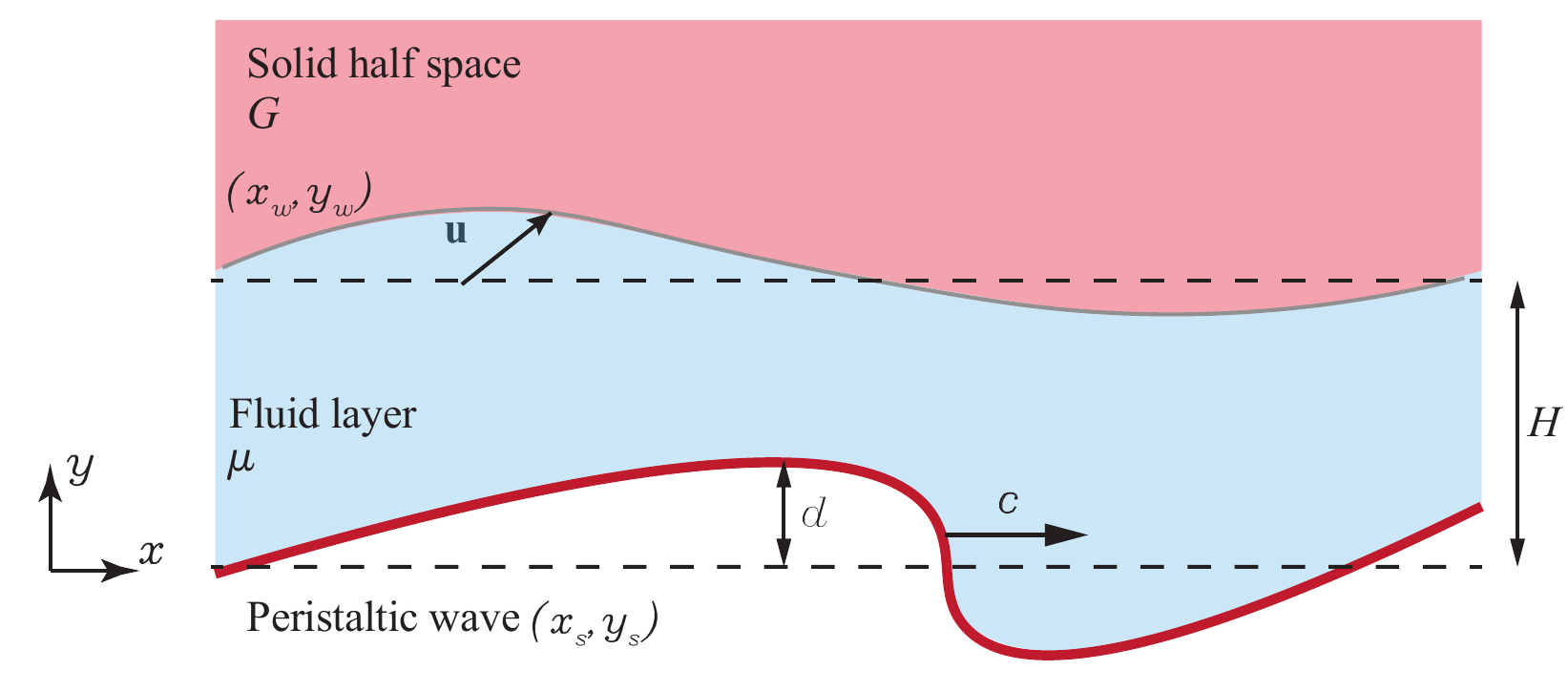}
    \caption{A wave of small amplitude and long wavelength travels in the positive $x$-direction along the bottom boundary. A fluid layer of average thickness $H$ has viscosity $\mu$. A linear elastic half space is located at $y=H$ and has elastic shear modulus $G$. The coordinates $(x_s,y_s)$ and $(x_w,y_w)$ describe the driving wave and elastic interface, respectively. The diagram is not to scale. }
    \label{fig:schematic}
\end{figure*}
We develop a model for peristalsis by re-framing the analysis of \cite{taylor_1951} for the swimming velocity of an undulating sheet through an unbound viscous fluid. A fixed sheet carries a train of waves which drive flow near an elastically deforming half space. The manuscript is organized as follows. In Section 2, an analytical solution to the fluid flow is found through an asymptotic expansion of boundary conditions assuming a small peristaltic amplitude relative to the wavelength. Additionally, the elastic motion of the solid half space is determined as a result of the oscillatory motion of the fluid. In Section 3, we compare two orthogonal peristaltic wave motions, namely, vertically oscillating transverse waves and horizontally oscillating longitudinal waves. We quantify the induced flow rate and elastic deformation in each case. Special attention is paid to how the flow and deformation are dependent on the dimensionless stiffness of the half space. The deformation into the solid half space is determined as well as phase differences between the driving wave and elastic surface waves. We highlight biologically relevant quantities as they pertain to the glymphatic system in Appendix B. In this case, we reduce an axisymmetric annulus to a 2D half space. The peristaltic wave travelling along the lower boundary is recast as the wave-like displacement of the arterial wall, the viscous fluid layer as CSF in the perivascular space and the linear elastic solid as soft brain tissue. We note the similarity of our work with the recent study of \cite{jha2024taylorsswimmingsheetnear} but highlight several distinctions. We consider a semi-infinite elastic medium, accounting for both vertical and horizontal deformations and thus determine two distinct elastic surface waves which arise from transverse and longitudinal peristaltic motion.
\section{Model Description}
\subsection{Governing equations}
In a rectangular domain, shown in Fig.~\ref{fig:schematic}, a fluid layer with viscosity $\mu$ and average thickness $H$ flows in the space between a peristaltic wave and an elastic half space. The Reynolds number of the flow is assumed to be small, meaning $Re=\rho_fc/\mu k\ll1$, where $\rho_f$ is the fluid density. The fluid motion is described by the Stokes equations,
\begin{equation}
    \begin{aligned}
      &\nabla{P} = \mu\nabla^2{\mathbf{V}}, \\
      &\nabla\cdot\mathbf{V} = 0,
    \end{aligned}
\end{equation}
where $P$ and $\mathbf{V}$ are the pressure and velocity of the fluid domain, respectively. We take the solid half space to be linear elastic and govern the motion with the incompressible Navier-Cauchy equations
\begin{equation}
    \begin{aligned}
    &\nabla{p} = G\nabla^2{\mathbf{u}},\\
    &\nabla\cdot\mathbf{u} = 0,
        \end{aligned}
\end{equation}
where $p$, $G$ and $\mathbf{u}$ are the pressure, shear modulus and displacement field in the elastic domain, respectively. The peristaltic wave, at coordinates $(x_s,y_s)$, travels in the positive $x$-direction and its motion is prescribed by  
  \begin{equation}
    \begin{aligned}
    &x_s(x, t)=x_0+b \sin [k(x-c t)], \\
    &y_s(x, t)=d_1 \sin[ k(x-c t)]+d_2 \sin [2 k(x-c t)],
    \end{aligned}
\end{equation}
where $b$ is the longitudinal wave amplitude and $d_1$ and $d_2$ the amplitudes of the first and second transverse modes, respectively. The wave number and speed are $k$ and $c$, respectively. The prescribed motion of the peristaltic wave induces fluid flow which in turn deforms the elastic half space. In biological peristaltic pumping, the driving wave motion is often not sinusoidal. For example, longitudinal peristalsis is found in the ureter and non-sinusoidal wave shapes are present in peristaltic pumping of cerebrospinal fluid. Thus, we consider a higher-order transverse wave mode and a longitudinal wave to study the effect of wave shape on the fluid flow and solid deformation properties. The shape of the interface between the fluid and solid is defined as
\begin{equation}
    (x_w,y_w) = (x,H) + \mathbf{u}(x,H,t).
\end{equation}
The form of the governing equations for $\mathbf{V}$ and $\mathbf{u}$ leads to a convenient set of general solutions. By choosing the appropriate form of solution to the biharmonic stream function for both the fluid and solid, incompressibility is automatically satisfied for each. The task is to apply the no-slip and force balance boundary conditions dictated by the geometry, thus solving for the velocity and displacement fields.

\subsection{Boundary Conditions}
The boundary conditions imposed in the problem introduce nonlinearity. Therefore, the conditions are expanded in Taylor series about $y=0$ and $y=H$ in orders of $bk,  d_1k$ and $d_2k$, where it is assumed that the peristaltic amplitudes are small relative to the wave number. The problem is then linearized and solved asymptotically. The interface deformations are found at first order while an averaged nonzero flow rate is obtained at second order. On the peristaltic wave, the no slip condition requires the fluid velocity be that of the oscillating boundary:
\begin{equation}
    \begin{aligned}
            &V_x(x_s,y_s)  = \frac{\partial{x_s}}{\partial{t}} 
             = -bk \cos [k(x-c t)], \\
            &V_y(x_s,y_s) =  \frac{\partial{y_s}}{\partial{t}} 
            =  -d_1 k \cos [k(x-c t)] -2d_2 k \cos [2k(x-c t)].
    \end{aligned}
    \label{eq:bc1}
\end{equation}
We expand the left hand side of these equations about $y=0$ to obtain
\begin{equation}
    \begin{aligned}
        \mathbf{V}(x_s,y_s) &= \bigg(\frac{\partial{x_s}}{\partial{t}},\frac{\partial{y_s}}{\partial{t}}\bigg)\\ 
        &= \mathbf{V}^{(1)}(x,0) + y_s\frac{\partial{\mathbf{V}}^{(1)}}{\partial{y}}\bigg|_{(x,0)}
        + x_s\frac{\partial{\mathbf{V}}^{(1)}}{\partial{x}} \bigg|_{(x,0)} \cdots,
    \end{aligned}
\end{equation}
where superscripts indicate the order of magnitude in the wave amplitude.
At the fluid-solid interface, the no slip condition requires the fluid velocity match the velocity of the deforming solid, i.e,
\begin{equation}
    \mathbf{V}(x_w,y_w) = \frac{d}{d{t}}\mathbf{u}(x_w,y_w,t),
     \label{eq:bc2}
\end{equation}
thus,
\begin{equation}
     \mathbf{V}\big((x,H) + \mathbf{u}(x,H,t)\big)  = \frac{d}{dt}\mathbf{u}\big((x,H) + \mathbf{u}(x,H,t)\big).
\end{equation}
The left hand side of Eq.~(2.8) must be expanded about $y=H$ to obtain 
\begin{equation}
        \mathbf{V}\big((x,H) + \mathbf{u}(x,H,t)\big) 
        = \mathbf{V}^{(1)}(x,H) + \mathbf{u}^{(1)}(x,H,t)\cdot\nabla\mathbf{V}^{(1)}(x,H) \cdots
    \label{eq:VexpBC2}
\end{equation}
and the right hand side is expanded about $y=H$ to obtain
\begin{equation}
    \begin{aligned}
        \frac{d}{dt}\mathbf{u}\big((x,H,t) + \mathbf{u}(x,H,t)\big) &= \frac{\partial}{\partial{t}}\mathbf{u}^{(1)}(x,H,t) \\
        &+ \frac{d}{d{t}}\mathbf{u}^{(1)}(x,H,t)\cdot\nabla\mathbf{u}^{(1)}(x,H,t) \cdots.
     \label{eq:uexpBC2}
     \end{aligned}
\end{equation}
 We equate the right hand sides of Eqs.~(\ref{eq:VexpBC2}) and (\ref{eq:uexpBC2}) and match orders of magnitude. The force balance at the interface will determine the shape of $(x_w,y_w)$, and thus the motion within the elastic half space, $\mathbf{u}(x,y,t)$. The normal and tangential components of the stress balance at the wall are
\begin{equation}
    \begin{aligned}
        &\bigg[p +2G\frac{\partial u_y}{\partial y} = P + 2\mu\frac{\partial V_y}{\partial y} \bigg]_{(x_w,y_w)},\\
        &\bigg[G\bigg(\frac{\partial u_x}{\partial y}+\frac{\partial u_y}{\partial x}\bigg) = \mu\bigg(\frac{\partial V_x}{\partial y}+\frac{\partial V_y}{\partial x}\bigg)\bigg]_{(x_w,y_w)},
    \end{aligned}
     \label{eq:bc3}
\end{equation}
respectively. 
\subsubsection{Nondimensionalization}
The above equations are nondimensionalized by considering $1/k$ and $1/ck$ as the characteristic length and time scales, respectively. The following dimensionless parameters are defined: $\Tilde{H}=Hk, \Tilde{b}=bk, \Tilde{d_1}=d_1k, \Tilde{d_2}=d_2k, \Tilde{t}=tck, \Tilde{x}=xk,$ and $\Tilde{y}=yk$. 

The dimensionless parameter $\mathit{\Lambda} = G/{\mu ck}$ is naturally arises from nondimensionalizing the stress balance at the wall (Eq. 2.11). As $\mathit{\Lambda}\rightarrow\infty$, the upper domain is rigid, and as $\mathit{\Lambda}\rightarrow 0$ the upper domain is highly deforming. It should be noted that this number is proportional to $Re/Ca$, where $Re = \rho_f c/\mu k$ is the Reynolds number and $Ca = \rho_s c^2/G$ is the Cauchy number.
\subsection{Solution}
We seek a solution to the velocity field to obtain a flow rate through the channel as a function of physical parameters. Due to the oscillatory nature of the flow, we specify the average flow rate over a period as,
\begin{equation}
    Q  =  \frac{1}{2 \pi}\int_{0}^{2 \pi}\int_{0}^{\Tilde{H}} \Tilde{V_x}\,d\Tilde{y} \,d\Tilde{t}.
\end{equation}
Since the motion is incompressible, it is convenient to use the stream functions $\psi_f$ and  $\psi_s$, where $\mathbf{V}=\nabla \times \psi_f\hat{\mathbf{z}}$ and $\mathbf{u}=\nabla \times \psi_s\hat{\mathbf{z}}$. The governing equations then imply $\nabla^4\psi_f=0$ and $\nabla^4\psi_s=0$. The geometric constraints of the problem guide our choice of the general solutions. Since the fluid layer is confined, the solution will contain hyperbolic trigonometric functions. The elastic solid is taken to be semi-infinite and thus the deformations are expected to decay exponentially with the wavelength. We choose functions that satisfy these conditions to obtain general solutions of the form,

\begin{equation}
    \begin{aligned}
\psi_f^{(n)}= & \sum_{m=1}^{\infty}\left\{\left[\left(A_{n, m}^{(f)}+E_{n, m}^{(f)} \Tilde{y}\right) \cos m(\Tilde{x}-\Tilde{t})\right.\right. \\
& \left.+\left(B_{n, m}^{(f)}+F_{n, m}^{(f)} \Tilde{y}\right) \sin m(\Tilde{x}-\Tilde{t})\right] \cosh m \Tilde{y} \\
& +\left[\left(C_{n, m}^{(f)}+G_{n, m}^{(f)} \Tilde{y}\right) \cos m(\Tilde{x}-\Tilde{t})\right. \\
& \left.\left.+\left(D_{n, m}^{(f)}+H_{n, m}^{(f)} \Tilde{y}\right) \sin m(\Tilde{x}-\Tilde{t})\right] \sinh m \Tilde{y}\right\} \\
& +\alpha_n \Tilde{y}+\beta_n \Tilde{y}^2+\gamma_n \Tilde{y}^3
\\
\psi_s^{(n)}=  &\sum_{m=1}^{\infty}\left[\left(A_{n, m}^{(s)}+E_{n, m}^{(s)} \Tilde{y}\right) \cos m(\Tilde{x}-\Tilde{t})\right. \\
& \left.+\left(B_{n, m}^{(s)}+F_{n, m}^{(s)} \Tilde{y}\right) \sin m(\Tilde{x}-\Tilde{t})\right] e^{-m \Tilde{y}},
\end{aligned}
    \label{eq:stream}
\end{equation}
for order $n$ and mode $m$. The flow and deformation are determined by solving for the coefficients in Eq. (\ref{eq:stream}) by applying the boundary conditions in Eqs. (\ref{eq:bc1}), (\ref{eq:bc2}), and (\ref{eq:bc3}).

\section{Results}\label{sec:types_paper}
\subsection{Solid deformation}

\begin{figure}
    \centering
    \includegraphics[width=1\linewidth]{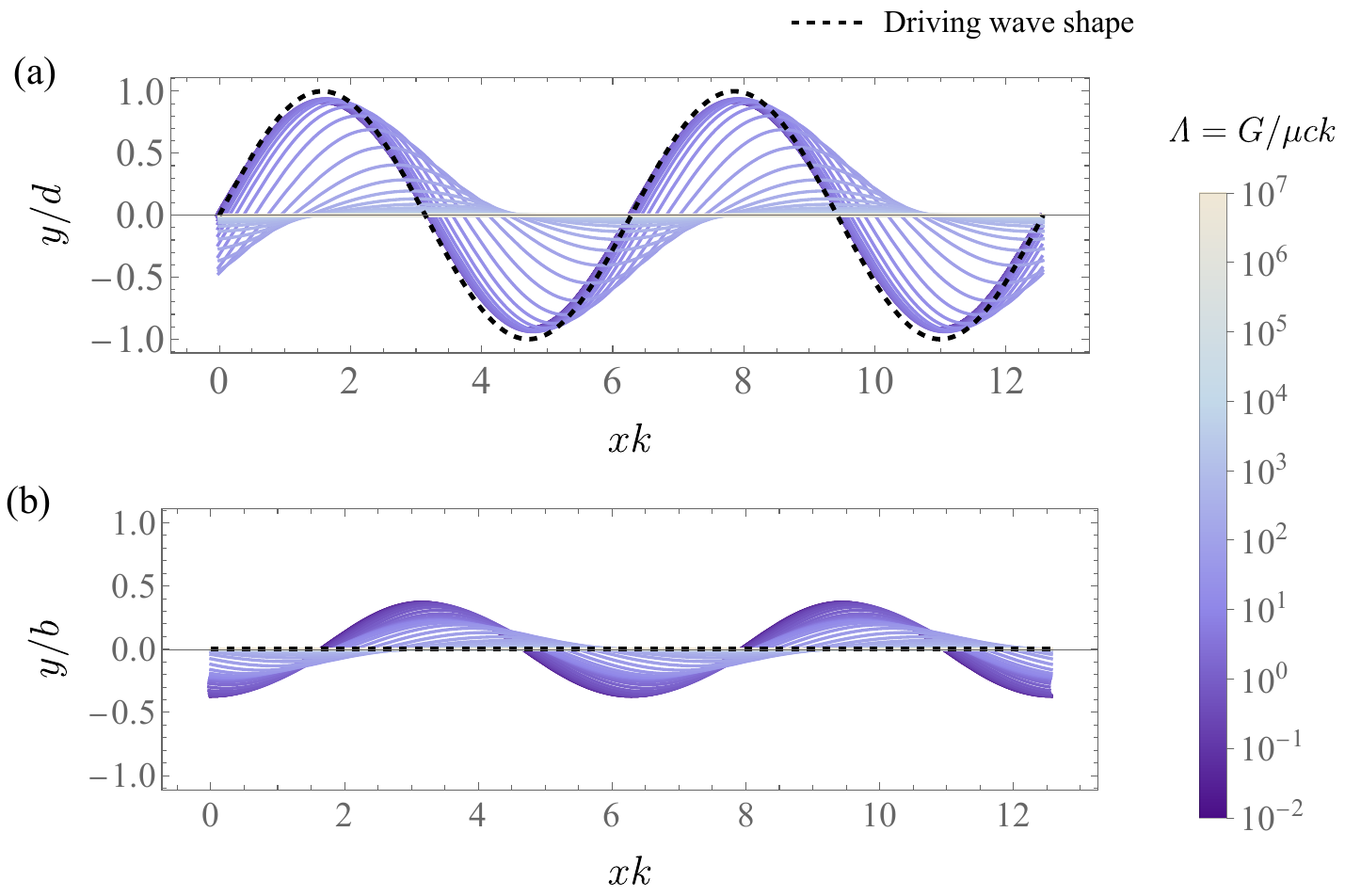}
    \caption{The shape of the interface between the elastic half-space and fluid channel for (a) transverse and (b) longitudinal driving waves of equal wavelength, speed, amplitude, and phase over a large range of $\mathit{\Lambda}$. The deformation amplitude is normalized by the wave amplitude of the lower wall, denoted by $d$ and $b$ in (a) and (b), respectively. Here $Hk = 0.5$ and $dk=bk=0.1$.}
    \label{fig:deform0}
\end{figure}

Although there is no time-averaged flow rate to first order in the amplitude of the oscillating boundary, the oscillatory flow within the fluid layer produces deformation in the elastic half space. We consider the interface shape over a range of $\mathit{\Lambda}$ values and then decompose the interface deformation into horizontal and vertical components. Figure \ref{fig:deform0} shows the interface shape produced from each case over a range of the stiffness parameter, $\mathit{\Lambda}$, for fixed $H$. In each case, the deformations are periodic of equal wavelength and speed to the driving peristaltic wave. A transverse driving wave produces relatively large vertical deformations at the interface. As expected, deflections are largest for low stiffness where the elastic solid deforms easily from fluid interaction. For $\mathit{\Lambda}$ approximately $O(10^{-1})$ or less, the interface shape is in phase with the driving wave, shown as a dotted line. As stiffness increases, the amplitude of vertical deformation decreases and the phase difference between the interface and driving wave plateaus to $\pi/2$.

 A longitudinal driving wave will produce smaller vertical deformations of the elastic interface as compared to a transverse wave of equal amplitude. While vertical deflections are still larger in the small $\mathit{\Lambda}$ limit, contrary to the transverse case, the resultant interface shape is out of phase with the driving wave for small $\mathit{\Lambda}$. Here, the phase difference between the driving and interface waves reaches $\pi$ as stiffness increases. Intuitively, the transverse wave produces higher normal forces to the fluid-elastic interface as compared to the longitudinal wave which produces a higher amount of shear stress. Thus, vertical displacement is significantly larger in the transverse case whereas horizontal displacements are greater in the longitudinal case. Between the two cases, this difference in the direction of forces at the interface causes the difference in the deformation phases.

\begin{figure}
    \centering
    \includegraphics[width=1\linewidth]{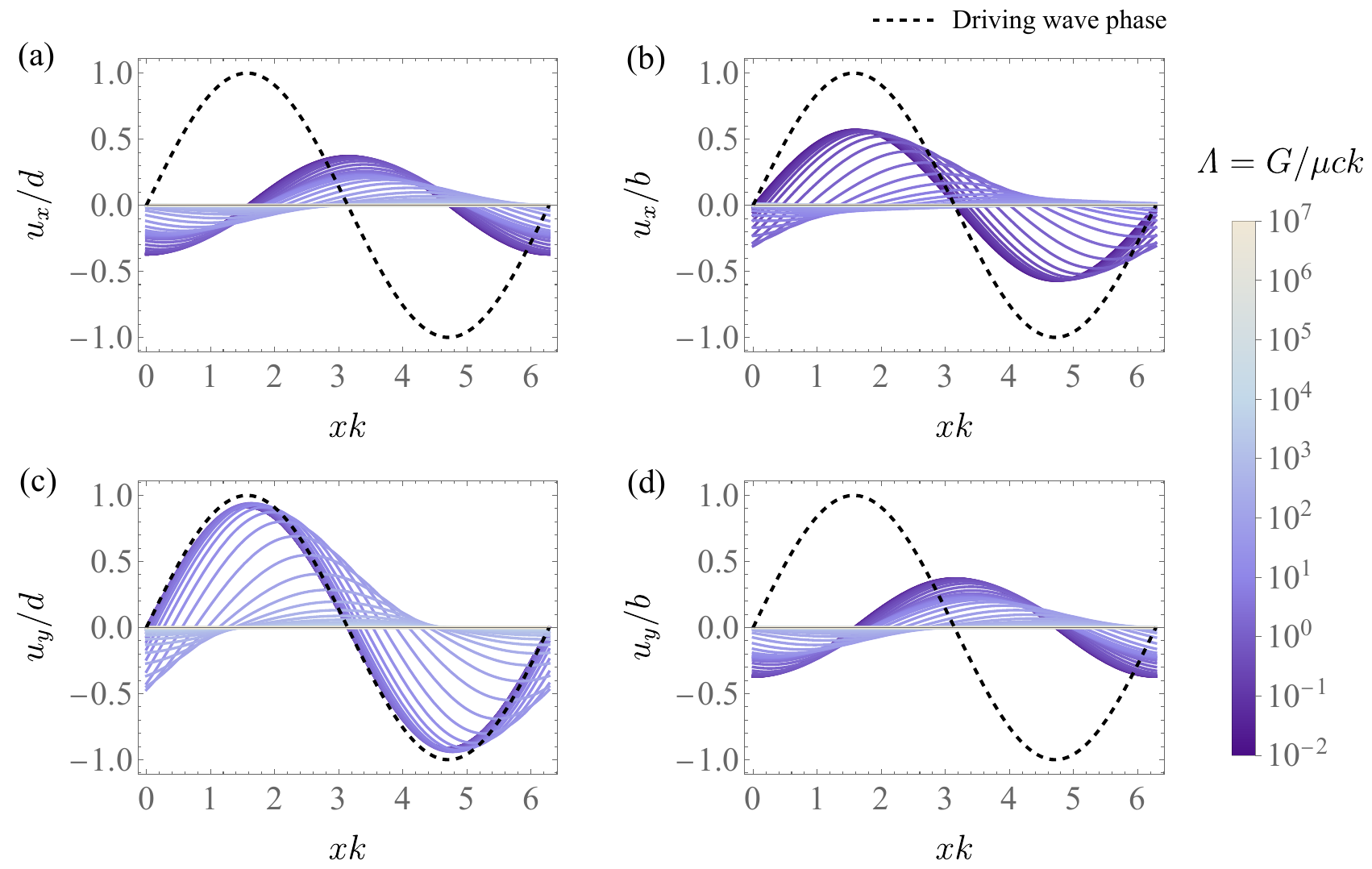}
    \caption{Horizontal (a,b) and vertical (c,d) deformations at the interface for $Hk = 0.5$ over a period of the driving peristaltic profile over a range of stiffness parameter, $\mathit{\Lambda}$, for transverse (a,c) and longitudinal (b,d) driving peristaltic waves.}
    \label{fig:Ud}
\end{figure}

 To further illustrate this point, we look to the components of deformation at the interface. The vertical, $u_y$, and horizontal,$u_x$, deformations at the interface are plotted over a period of the driving wave in Figure~\ref{fig:Ud}. Deformations resulting from a transverse peristaltic wave are normalized by $d_1=d$ and deformations from a longitudinal wave are normalized by $b$. The phase of $u_x~(u_y)$ of the transverse driven deformation matches the phase of $u_y~(u_x)$ of the longitudinal driven deformation. The phase of $u_x$ in the longitudinal case is relatively more dependent on $H$ as compared to the other delfections. For this component, as $Hk$ increases, the deformation moves out of phase of the driving wave, the deformations in the soft limit eventually plateau to be antiphase of the peristaltic wave at a dimensionless height of $Hk=0.5$.

\begin{figure}
    \centering
    \includegraphics[width=1\linewidth]{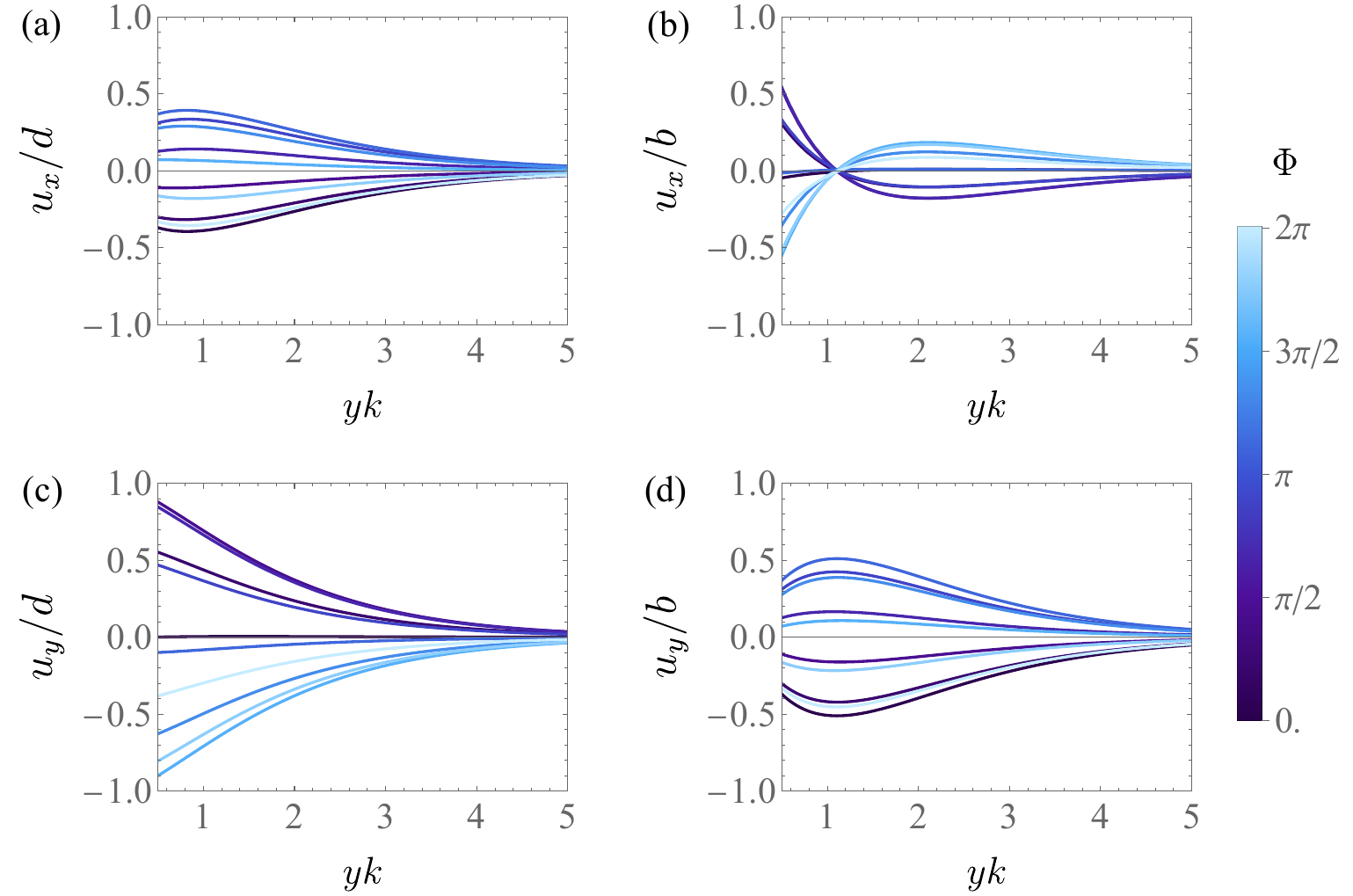}
    \caption{Elastic deformation for $Hk=0.5$ as a function of $yk$ (for $yk>0.5$), the depth into the elastic material where curves represent displacement at a certain moment in the peristaltic wave period, $\Phi$. Horizontal and vertical displacement as a result of a transverse driving wave (a and c respectively) and as a result of a longitudinal driving wave  (b and d respectively). Displacements are normalized by the respective driving wave amplitude.}
    \label{fig:DeformDecay}
\end{figure}

 We next consider the motion of elastic material into the half space in Figure~\ref{fig:DeformDecay}. A transverse driving wave leads to prograde orbits of material points, regardless of the depth into the solid. When the deformation is plotted as a function of $y$, (see Fig. \ref{fig:DeformDecay}a,c), we see that the amplitudes decay into the elastic medium. 
When the system is driven with a longitudinal peristaltic profile, orbits at the surface are in retrograde to the direction of the travelling wave, while orbits at a depth greater than $1/5k$ are in prograde. This is evident in Fig.~\ref{fig:DeformDecay}(b) as curves cross the $x$-axis. This type of elastic motion is similar qualitatively to what is observed in Rayleigh surface waves \cite{Ray}, though we note the following distinctions. The waves seen here are quasi-static and nondispersive. Further, they are continually driven by traction forces over the surface rather than from an initial, transient excitation. The two driving wave types induce two kinds of elastic surface waves which are differentiated by the phase difference between vertical and horizontal deformations. This is a result of the direction of the dominant traction forces present at the elastic interface induced by the peristaltic wave.

\subsection{Induced flow}

\begin{figure}[h]
    \centering
    \includegraphics[width=1\linewidth]{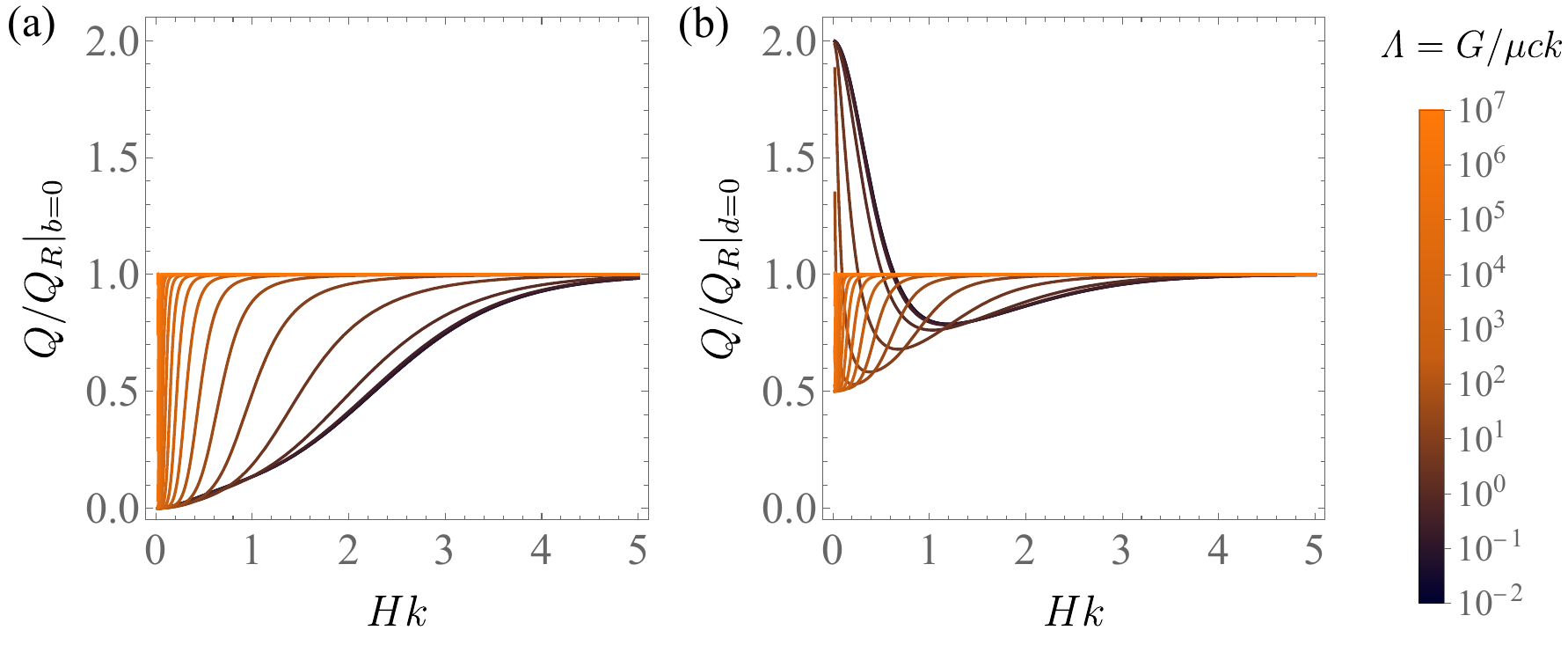}
    \caption{Average flow rate, $Q$, normalized by the flow rate near a perfectly rigid half-space, $Q_R = Q(\mathit{\Lambda} \rightarrow \infty)$. For a longitudinal peristaltic wave (a), it is found that half-space flexibility decreases flow at large $H$ but increases flow at small $H$. For a transverse driving wave (b), it is found that flexibility of the elastic half space decreases the average flow rate. Qualitatively, this is due to momentum being used in the $y$-direction to deform the surface rather than to push fluid in the $x$-direction.}
    \label{fig:QbdVsH}
\end{figure}

Nonzero average flow occurs at second order in the amplitude of the oscillating boundary. The coefficients to $Q$ are shown in Appendix A. To analyze the flow, we compare $Q$ with the flow rate found for peristaltic pumping near a rigid boundary \cite{katz_propulsion_1974}:
\begin{equation}
    \begin{aligned}
Q_R=\lim_{\mathit{\Lambda} \to\infty} Q  =\frac{1}{4}H\left[-b^2+ d_1^2 \left(1+\frac{4 H^2}{-2 H^2+\cosh (2 H)-1}\right)\right].
    \end{aligned}
\end{equation}
To determine the effects of the elastic boundary on the flow, we plot $Q$ versus $H$ over a range of the stiffness parameter $\mathit{\Lambda}$ (see Figure \ref{fig:QbdVsH}). We normalize $Q$ by the flow near a rigid boundary, $Q_R$. Figure~\ref{fig:QbdVsH}b presents flow results for a longitudinal driving peristaltic wave. For small $H$ and small $\mathit{\Lambda}$, flow is increased as compared to flow near a rigid boundary. This is due to the fact that when the fluid thickness is small and the elastic material is soft, the interface will propagate longitudinal waves, with minimal vertical displacement, in the same direction as the driving wave. A second peristaltic wave is produced which doubles the flow rate, hence the collapse onto $Q/Q_R=2$ in the small $H$ limit. As $H$ grows, this effect lessens and flow reduces before matching the rigid limit for large $H$. Moreover, as $H$ grows, the vertical deformation at the interface grows relative to the horizontal deformation. Thus, momentum is being transferred in the vertical direction, and does not contribute to the time averaged flow rate. 

In the transverse case, small values of $H$ and $\mathit{\Lambda}$ result in little to no flow as compared to large $\mathit{\Lambda}$ (see Fig.~\ref{fig:QbdVsH}a), directly contrasting results of the longitudinal wave. The curves collapse onto the rigid limit at large $H$. Curves associated with small $\mathit{\Lambda}$, in darker colors, approach the rigid limit at larger $H$ than lighter curves. This behavior is expected as a softer interface must be positioned relatively far away from the oscillating boundary to not be mechanically disturbed as compared to a harder interface.  We find that having a passively deformable boundary results in less flow compared to a rigid boundary for a transverse peristaltic wave. This is due to motion from the driving wave being transferred vertically into the deforming the solid. In the rigid case, momentum from the driving wave enters the fluid only. When near a deformable boundary, momentum deforms the elastic solid. Though it is linearly elastic and thus no momentum is diffused upward into the solid, energy is spent in the vertical direction which does not contribute to the horizontal flow of fluid and therefore the time averaged flow rate. Qualitatively, streamlines of the fluid flow have cyclic, recirculating paths near a soft boundary whereas streamlines have nearly sinusoidal meandering paths when close to a rigid boundary.

\begin{figure}[h]
    \centering
    \includegraphics[width=1\linewidth]{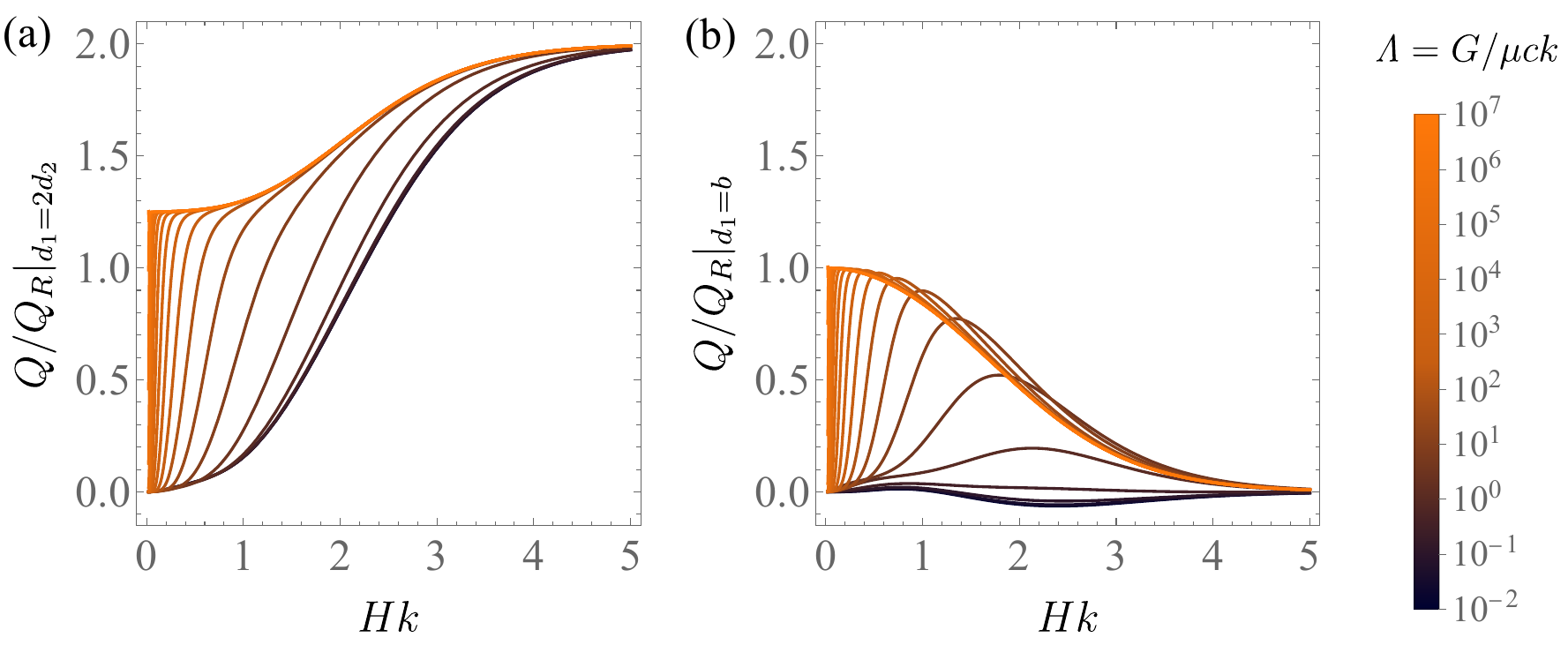}
    \caption{Flow for a system driven by both transverse and longitudinal waves of equal magnitude (a) and by two modes of a transverse wave (b). A longitudinal wave opposes the transverse wave and creates back flow, whereas an additional transverse mode increases flow.}
    \label{fig:flowcompare}
\end{figure}

Figure \ref{fig:flowcompare}  presents results from linear combinations of peristaltic waves. In Fig.~\ref{fig:flowcompare}a, flow is plotted for a combination of a transverse and longitudinal wave $b=d_1=0.1, d_2=0$ and in Fig.~\ref{fig:flowcompare}b is two modes of transverse waves, $d_1=2d_2=0.1, b=0$. A longitudinal peristaltic wave will produce flow in the opposite direction in which it is travelling, a flow known as reflux. This is in agreement with results of previous studies \cite{katz_propulsion_1974,reynolds_swimming_1965,shaik_swimming_2019, Blake_1971}, yet contradicts the results of \cite{kalayeh_longitudinal_2023} who found longitudinal motion of the peristaltic wall limits reflux. In the case of $b=d_1$ and $d_2=0$, shown here in Fig. \ref{fig:flowcompare}a, the transverse and longitudinal components of the wave balance each other, and flow decays to zero at large $H$. At small $H$, flow rates with large values of $\mathit{\Lambda}$ are comparable to $Q_R$ while flow rates with small $\mathit{\Lambda}$ are near zero. As $H$ grows, curves representing high stiffness decay to zero while curves of low stiffness ($\mathit{\Lambda} \approx 1$) fluid travels in the negative $x$-direction before approaching back to zero. Contrary to flow rates with $b=0$, for $b \neq 0$ flow rates near softer interface can exceed that of flow rate near stiff interfaces, shown by the darker curves which surpass bright orange curves in the left panel. For $b > d$, the horizontal asymptote reached at large $H$ becomes negative while for $b < d$ the asymptote is positive. The larger the magnitude of $b$, the further down this limit is shifted, resulting in increased reverse flow.

The resultant flow from considering higher modes of transverse waves is simply the linear combination of two flow solutions due to the linearity of the problem. An additional mode will contribute a corresponding amount of additional flow, as seen in Figure \ref{fig:flowcompare}b. Although we calculate the time averaged flow rate, which neglects the phase difference between the two modes, the work of \cite{bauerle_living_2020} suggests this phase difference could be optimized such that the amplitudes of the two modes interact constructively, thus maximizing flow.

\section*{Conclusions}
Peristaltic flow driven by a prescribed wave form through an elastic channel is solved analytically. The deformation of the elastic half space is determined and analyzed in two distinct cases, transverse and longitudinal driving waves. It is found that these two means of driving the system will produce differing material motion of the elastic solid half space. A transverse peristaltic wave produces relatively large vertical deformations of the elastic interface which decay monotonically into the elastic medium. Conversely, a longitudinal peristaltic wave produces relatively large horizontal deformations at the interface which do no decay monotonically into the medium but rather follow a similar trend to the motion observed in Rayleigh surface waves. Fluid flow rate is calculated through the channel in each case and compared over a range of the stiffness parameter. Transverse waves drive fluid forward, pushing fluid along in its direction of travel, whereas longitudinal waves drive fluid backward, pulling fluid against the direction of travel, in a mechanism known as reflux. It is found that decreasing stiffness of the upper elastic half space decreases overall flow rate of fluid through the channel for a transverse peristaltic wave regardless of the channel thickness. However, for a thin channel, low stiffness of the elastic material can increase flow rate when compared to a stiffer material for longitudinal peristaltic waves. We present our model in a generalized framework as well as within the context of perivascular pumping of cerebrospinal fluid in the brain, detailed in Appendix B. The analysis suggests it is plausible that perivascular pumping is a driver of CSF flow in annular perivascular spaces. Future work will consider a nonlinearly elastic or viscoelastic solid half space. Alternatively, implementing porosity to the solid half space could be informative particularly in the study of flows near biological tissue.

\appendix
\section{Expressions for constants which appear in general solutions}
Expressions for the constants found in the general solutions to the biharmonic equations in Eqs. \ref{eq:stream} are provided. Although solutions to both longitudinal and transverse waves, as well as higher modes, are considered in this paper, the expressions presented here are the single mode case of a transverse wave for brevity.
\subsection{$O(d^1)$ constants}

\begin{equation}
    \begin{aligned}
&A^{f}_{(1,1)}= 0,\\
    \end{aligned}
\end{equation}
\begin{equation}
    \begin{aligned}
&C^{f}_{(1,1)}= \frac{4 d\mathit{\Lambda}  \left(\mathit{\Lambda} ^2-\left(\mathit{\Lambda} ^2+1\right) \left(\left(2 H^2+1\right) \cosh (2 H)+2 H \sinh (2 H)\right)-1\right)}{k \mathcal{D}},\\
            \end{aligned}
\end{equation}
\begin{equation}
    \begin{aligned}
&E^{f}_{(1,1)}= \frac{4 d \mathit{\Lambda}  \left(-\mathit{\Lambda} ^2+\left(\mathit{\Lambda} ^2+1\right) \left(\left(2 H^2+1\right) \cosh (2 H)+2 H \sinh (2 H)\right)+1\right)}{\mathcal{D}},\\
            \end{aligned}
\end{equation}
\begin{equation}
    \begin{aligned}
&G^{f}_{(1,1)}= -\frac{8 d H^2 \mathit{\Lambda}  \left(\mathit{\Lambda} ^2+1\right) \sinh (2 H)}{\mathcal{D}},\\
            \end{aligned}
\end{equation}
\begin{equation}
    \begin{aligned}
&B^{f}_{(1,1)}= \frac{d}{k},\\
            \end{aligned}
\end{equation}
\begin{equation}
    \begin{aligned}
&D^{f}_{(1,1)}= \frac{d \mathcal{C}}{k \mathcal{D}},\\
            \end{aligned}
\end{equation}
\begin{equation}
    \begin{aligned}
&F^{f}_{(1,1)}= \frac{-d \mathcal{C}}{\mathcal{D}},\\
            \end{aligned}
\end{equation}
\begin{equation}
    \begin{aligned}
&H^{f}_{(1,1)}= \frac{
    \begin{bmatrix} 
        d \bigg(-3 \mathit{\Lambda} ^4+2 \mathit{\Lambda} ^2-4 H^2 \left(\mathit{\Lambda} ^2+1\right)^2
        +4 \left(H^2+1\right) \left(\mathit{\Lambda} ^4-1\right) \cosh (2 H)\\
        -\left(\mathit{\Lambda} ^2+1\right)^2 \cosh (4 H)-3\bigg)\end{bmatrix}}{\mathcal{D}},\\
                    \end{aligned}
\end{equation}
\begin{equation}
    \begin{aligned}
&A^{s}_{(1,1)}= \frac{
    \begin{bmatrix}
       16 d e^{2 H} \cosh ^4(H)\bigg(e^{6 H} \left(\mathit{\Lambda} ^2+1\right)+(2 (H-1) H+1) \left(\mathit{\Lambda} ^2+1\right)\\
       +e^{2 H} \left(-\mathit{\Lambda} ^2+4 H (2 H ((H-1) H+1)-1) \left(\mathit{\Lambda} ^2+1\right)+3\right)\\
       -e^{4 H} \left(\mathit{\Lambda} ^2+2 H \left((H-3) \mathit{\Lambda} ^2-3 H+1\right)-3\right)\bigg) 
    \end{bmatrix}
    }{\left(1+e^{2 H}\right)^4 k \mathcal{D}},\\
                \end{aligned}
\end{equation}
    \begin{equation}
    \begin{aligned}
&E^{s}_{(1,1)}\to \frac{
    \begin{bmatrix}
    16 d e^{2 H} \cosh ^4(H) \bigg(e^{6 H} \left(\mathit{\Lambda} ^2+1\right)-(2 H-1) \left(\mathit{\Lambda} ^2+1\right)\\
    +e^{4 H} \left(4 H^2-2 H+\left(4 H^2+6 H-1\right) \mathit{\Lambda}^2+3\right)\\
    -e^{2 H} \left(\mathit{\Lambda} ^2+4 H (H (2 H-1)+1) \left(\mathit{\Lambda} ^2+1\right)-3\right)\bigg)\end{bmatrix}}
    {\left(1+e^{2 H}\right)^4\mathcal{D}},\\
                \end{aligned}
\end{equation}
    \begin{equation}
    \begin{aligned}
&B^{s}_{(1,1)}\to \frac{
    \begin{bmatrix}
    16 d e^{2 H} \mathit{\Lambda} \cosh ^4(H) \bigg(e^{6 H} \left(\mathit{\Lambda} ^2+1\right)-(2 (H-1) H+1) \left(\mathit{\Lambda} ^2+1\right)\\
    +e^{2 H} \left(3 \mathit{\Lambda} ^2+4 H (2 H ((H-1) H+1)-1) \left(\mathit{\Lambda} ^2+1\right)-1\right)\\
    +e^{4 H} \left(-3 \mathit{\Lambda} ^2+2 H \left(-3 H \mathit{\Lambda} ^2+\mathit{\Lambda} ^2+H-3\right)+1\right)\bigg) \end{bmatrix}}{\left(1+e^{2 H}\right)^4 k\mathcal{D}},\\
                \end{aligned}
\end{equation}
    \begin{equation}
    \begin{aligned}
&F^{s}_{(1,1)}\to \frac{
    \begin{bmatrix}
        16 d e^{2 H} \mathit{\Lambda}  \cosh ^4(H) \bigg(e^{6 H} \left(\mathit{\Lambda} ^2+1\right)+(2 H-1) \left(\mathit{\Lambda} ^2+1\right)\\
        +e^{2 H} \left(3 \mathit{\Lambda} ^2-4 H (H (2 H-1)+1) \left(\mathit{\Lambda} ^2+1\right)-1\right)\\
        +e^{4 H} \left(-3 \mathit{\Lambda} ^2+2 H \left(\mathit{\Lambda}^2-2 H \left(\mathit{\Lambda} ^2+1\right)-3\right)+1\right)\bigg)\end{bmatrix}}
        {\left(1+e^{2 H}\right)^4 \mathcal{D}},\\
            \end{aligned}
\end{equation}
where 
\begin{equation}
    \begin{aligned}
      &\mathcal{D} =  8 \left(\mathit{\Lambda} ^2+1\right)^2 H^4+8 \left(\mathit{\Lambda} ^2+1\right)^2 H^2+3 \mathit{\Lambda} ^4-2 \mathit{\Lambda} ^2\\
    &\hspace{20pt}-4 \left(2 H^2+1\right) \left(\mathit{\Lambda} ^4-1\right) \cosh (2 H)+\left(\mathit{\Lambda} ^2+1\right)^2 \cosh (4 H)+3,\\
    &\mathcal{C}=\left(8 H^3+4 H-\sinh (4 H)\right) \left(\mathit{\Lambda} ^2+1\right)^2-4 H \left(\mathit{\Lambda} ^4-1\right) \cosh (2 H)\\
    &\hspace{20pt}+2 \left(2 H^2+1\right) \left(\mathit{\Lambda} ^4-1\right) \sinh (2 H).\\
    \end{aligned}
\end{equation}

\subsection{$O(d^2)$ constants}
\begin{equation}
    \begin{aligned}
        &\alpha_2= \frac{d^2 \mathcal{G}}{2\mathcal{M}}, \beta_2= -\frac{d^2 \mathcal{G}}{4 H\mathcal{M}}, \gamma_2= 0,\\
        \end{aligned}
\end{equation}
where
\begin{equation}
    \begin{aligned}
        \mathcal{G} &= -8 H^4 \left(\mathit{\Lambda} ^2+1\right)^2-4 \left(\mathit{\Lambda} ^4-1\right) \cosh (2 H)\\
        &+\left(\mathit{\Lambda} ^2+1\right)^2 \cosh (4 H)+3 \mathit{\Lambda} ^4-2 \mathit{\Lambda} ^2+3\\
        \mathcal{M} &= 8 H^4 \left(\mathit{\Lambda} ^2+1\right)^2-4 \left(2 H^2+1\right) \left(\mathit{\Lambda} ^4-1\right) \cosh (2 H)\\
        &+8 H^2 \left(\mathit{\Lambda} ^2+1\right)^2
        +\left(\mathit{\Lambda} ^2+1\right)^2 \cosh (4 H)+3 \mathit{\Lambda} ^4-2 \mathit{\Lambda} ^2+3.
    \end{aligned}
\end{equation}
\section{Analysis of cerebrospinal fluid flow in perivascular spaces}
We use our model in the context of the flow of cerebrospinal fluid (CSF) through annular perivascular spaces (PVSs). By taking the PVS to be an axisymmetric space between the artery and brain, we cast our 2D model as the upper half of the length-wise cross section of an annulus. In this configuration, the peristaltic wave acts as the artery deforming periodically due to the heartbeat. It is this deformation which is hypothesized to push CSF along the PVS \cite{romano_peristaltic_2020,hadaczek_perivascular_2006,thomas_fluid_2019,wang_fluid_2011,Mestre2018,Mestre2020,kelley_cerebrospinal_nodate}. The elastic half space of our model is cast as brain tissue which deforms as a result of the oscillatory flow. 

The biologically relevant parameters are the CSF viscosity, brain shear modulus, PVS layer thickness, artery deformation wavelength, amplitude and pulse wave speed. CSF is widely regarded as being mostly water \cite{kelley_cerebrospinal_nodate} so here we take $\mu=0.001$ Pa$\cdot~$s. Experimental data has characterized brain mechanical properties over a range of magnitudes but here we take $G=1000$ Pa \cite{budday2017mechanical} to be the shear modulus of brain tissue. Geometric quantities of the system also vary greatly in the literature. PVS layer thickness has been reported on orders of $H \in [10^{-5},10^{-2}]$m and artery deformations are reported to have amplitudes of order $d \in [10^{-7}, 10^{-5}]$m , wavelengths of $\lambda \in [10^{-3}, 1]$m and wave speeds of $c \in [10^{-2}, 1]$m s$^{-1}$ \cite{romano_peristaltic_2020}. 
We find that $Hk\ll1$ which implies the use of lubrication theory. However, as shown in \cite{katz_propulsion_1974}, for this problem, lubrication theory and the biharmonic analysis agree well when the conditions $Hk\ll1$ and $bH\ll1$ are satisfied simultaneously. In the case of CSF flow, the wave amplitude is at least one order of magnitude less than the PVS thickness, thus satisfying the latter condition. Additionally, our biharmonic results for average velocity across the fluid layer agree well with the lubrication approximation for bacterial swimming over an elastic mucus reported by \cite{tchoufag_2019}. 

\begin{figure}
    \centering
    \includegraphics[width=.8\linewidth]{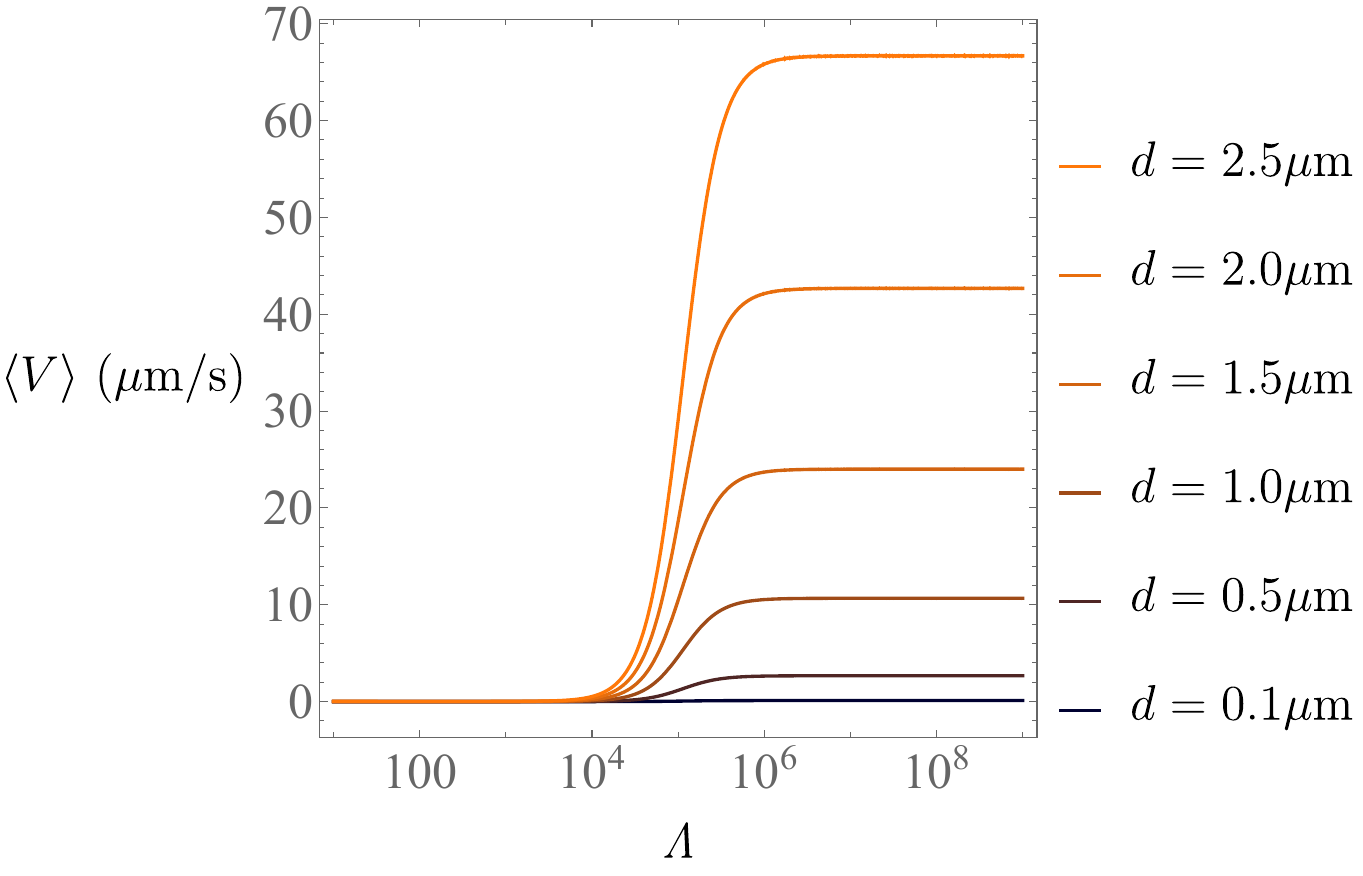}
    \caption{Average velocity, $\langle V \rangle$, across the perivascular space versus the softness parameter for a transverse peristaltic wave at six representative values of $d$, the peristaltic amplitude. Velocity is very nearly zero for all amplitude up until a critical $\mathit{\Lambda}$.}
    \label{fig:CSF_vel}
\end{figure}

Considering these biological quantities, we are now interested in values on the order of $\mathit{\Lambda} \in [10^{3},10^{7}]$, $Hk \in [10^{-4},10^{-2}]$ and $dk \in [10^{-5},10^{-3}]$. For best agreement between lubrication theory and the biharmonic analysis done here, we consider the largest scales of the geometry. We adopt values from the study conducted by \cite{romano_peristaltic_2020}. Namely, we choose $H=150\,\mu$m, $d$ to be on the order of $1\,\mu$m or less, and the peristaltic wavelength to be between $10^3$--$10^4\,\mu$m. We find an average flow speed through the PVS to be between the orders of 1--10\,$\mu$m/s, shown in Fig.~\ref{fig:CSF_vel}, which is consistent with reported experimental and theoretical data \cite{kelley_cerebrospinal_nodate}. We note that the velocity, like the flow rate, depends quadratically on the driving amplitude so small increases in $d$ produce large increases in the velocity. We find that no flow is achieved until a critical value of $\mathit{\Lambda}$ when velocity then quickly increases up to a plateau dependent on the driving amplitude. This observation agrees with the lubrication analysis of \cite{tchoufag_2019}.


\bibliography{Cilia,Peristalticflow,Poroelasticity,Reviews,Theory,lubrication,Experimental}

\end{document}